\documentclass[12pt]{article}
\usepackage{amsmath}
\DeclareMathOperator{\sech}{sech}
\newmuskip\pFqmuskip
\usepackage{subcaption}
\usepackage{textcomp}
\usepackage{cite}
\usepackage{float}
\usepackage[utf8]{inputenc}
\usepackage{amssymb}
\usepackage{graphicx}
\usepackage{scalerel,amssymb}

\begin{document}
		\begin{center}
		\vspace{0.4cm}{\large{\bf  The study of Kantowski-Sachs perfect fluid  cosmological model in modified gravity}}\\
		\vspace{0.4cm}{\it T. Vinutha$^{1}$,  K. Niharika$^{1}$ and K. Sri Kavya$^{2}$  }\\
		${}^{1}$Dept. of Applied Mathematics, AUCST, Andhra University, Visakhapatnam, India.\\
		${^2}$ Dept. of Mathematics, Maharaj Vijayaram Gajapathi Raj College of Engineering, Vizianagaram-535005, India.\\
		{$^{\star}$vinuthatummala@gmail.com }\\
	\end{center}
\begin{abstract}
	 Kantowski-Sachs perfect fluid cosmological model is explored in modified gravity with functional form $f(R,T)$=$f_1(R)$+$f_2(T)$  where $R$ is Ricci scalar and $T$ is the trace of energy-momentum tensor. With this functional form, three different cases have been formulated, namely negative and positive powers of curvature, logarithmic curvature and exponential curvature given by $f_1(R)=R+\gamma R^2-\frac{\mu^4}{R}$, $f_1(R)=R+\nu ln(\tau R)$ and $f_1(R)=R+\kappa e^{-\iota R}$ respectively, and for all these three cases, $f_2(T)=\lambda T$, here $\gamma$, $\lambda$, $\mu$, $\nu$, $\tau$, $\kappa$ and $\iota$ are constants. While solving the field equations, two constraints i)  Expansion scalar is proportional to  shear scalar ii) Hyperbolic scale factor are used. By using these conditions the required optimum solutions are obtained. The physical parameters are calculated and geometrical parameters of three cases are analysed against redshift($z$) with the help of pictorial representation. In the context of $f(R,T)$ gravity energy conditions are discussed with the help of pressure and energy density. If strong energy condition is positive the gravity should be attractive but in our model it shows negative it means that cosmic acceleration is due to antigravity, whereas NEC and DEC are fulfilled. The perturbation technique is used to test the stability of the background solutions of the obtained models. The inferences obtained from this paper are in persistent with the present cosmological observations and the model represents an accelerating universe. 
\end{abstract}

\textbf{Keywords}: Kantowski-Sachs spacetime, $f(R,T)$ theory, perfect fluid.\\ \indent\indent\indent\indent 

\section{Introduction} 
\indent Einstein's theory of general relativity is the foundation of modern physics and it describes black holes and gravitational phenomena but it break down to give an explanation of cosmic acceleration. In recent scenario it is well known that our universe is accelerating\cite{rie,per} and it is one of the trending topics in cosmology. To understand this mysterious concept, we focused on dark energy and modified theories of gravity. The universe is going through an accelerated period of expansion and it is revealed by the experiments such as CMBR and SN${I}a$. Dark energy can be inspected in many ways and reforming the geometric part of the Einstein-Hilbert action is regarded as the most efficient possible way and these changes lead to so many alternative theories of gravity. There are different classes of modified gravity such as  $f(R)$ gravity, $f(T)$ gravity, $f(G)$ gravity, $f(R,G)$ gravity. Among them  $f(R)$ gravity has attracted many researchers because it provides a natural gravitation alternative to dark energy. During the universe expansion $f(R)$ theory elucidate the change from deceleration phase to acceleration phase. $f(R)$ theory is presumed to be beneficial for resolution of the hierarchy problem or unification of grand unified theories with gravity in high energy physics. Nojiri and odintsov\cite{Nojiri:2010wj}, Nojiri et al.\cite{Nojiri:2017ncd}, Chatterjee and Jaryal\cite{chja}, Sotiriou and  Faraoni\cite{tof} and De Felice and Tsujikawa\cite{de} are some of the authors who worked on various cosmological models in $f(R)$ theories of gravity. A new class of $f(R,T)$ gravity presented by Harko et al.\cite{har}, by including trace $T$ in $f(R)$ theory. The $T$-dependence in $f(R,T)$ gravity may appear from the presence of imperfect fluids or quantum effects. Among all the modified theories of gravitation, the $f(R,T)$ theory is a generalized theory because there is an energy transfer relation between matter and geometry. The existence of this relationship is the cause of the rapid expansion of the universe. The authors who worked on $f(R,T)$  gravity are included in references \cite{noz,sha,sin,jam,corr,mom,sam,zus}.\\
\indent In this paper, we examine three specific cases one of them is combination of $\frac{1}{R^x}$ and $R^y$ i.e. $f(R)=R+\gamma R^y-\frac{\mu^4}{R^x}$ where $\gamma$ and $\mu$ are constants. In this functional form, it has both positive and negative curvature powers. At low curvature it leads to gravitational alternative for dark energy which helps in speed up of cosmic expansion where as high curvature describes the inflationary stage of early universe\cite{nod}. By considering $R^y$ term for $1<y<2$ power law inflation happen at early stage. If $y=2$, Starobinsky inflation takes place\cite{sta}, the term $R^2$ indicates natural correction to general relativity. According to Nojiri and Odinstov\cite{sov} $R^2$ term is necessary to get rid of instabilities, linear growth of the gravitational force, produce early time inflation and appear to pass the solar system tests. The state of no linear growth for gravitational force makes it very much fascinating. Higher derivative terms like $R^2$, $R^3$ can be used to put down the instabilities significantly. For equivalent scalar tensor theory the solar system test may be passed as scalar has large mass originated again by higher derivative terms. The standard Einstein's gravity may be modified by considering a $\frac{1}{R}$ term in the Einstein Hilbert action\cite{no} which represents the present acceleration of the universe. But the insertion of $\frac{1}{R}$ term  generates instabilities which can be overcome by addition of $R^2$ term to the Einstein's gravitational action. Besides the advantages of this functional form, have well acceptable Newtonian limit, no instabilities and no Brans Dicke problem in scalar tensor version. When we put $y=2$ and $x=1$ in the above functional form $f(R)=R+\gamma R^y-\frac{\mu^4}{R^x}$ it reduces to $f(R)=R+\gamma R^2-\frac{\mu^4}{R}$ and the obtained results are very efficient. In addition to this functional form by using the linear function of $f(T)=\lambda T$, we get the final form of $f(R,T)=R+\gamma R^2-\frac{\mu^4}{R}+\lambda T$ where $\gamma$, $\mu$ and $\lambda$ are constants. Vinutha et al. \cite{vv} have worked on Kantowski–Sachs perfect fluid cosmological model in $R^2$- Gravity. Vinutha and  Sri Kavya\cite{vink} have studied Bianchi type cosmological models in $f(R, T)$ theory with quadratic functional form. Brookfield\cite{bro} have worked on viability of $f(R)$ theories with additional powers of curvature. Godani and Samanta\cite{gos} have studied traversable warmholes on $f(R)$ gravity where $f(R)=R+\alpha R^n$. Banik et al.\cite{bai} have discussed Bianchi-I cosmological model in $f(R)=R-\frac{\beta}{R^n}$ gravity.\\
\indent Next, we consider logarithmic curvature i.e. $f(R,T)=R+\nu ln(\tau R)+\lambda T$ where $\tau$, $\nu$ and $\lambda$ are constants. As this modified gravity has put forward a gravitational alternative for dark energy, it is quite interesting to work on this particular functional form. In this model logarithmic  terms are produced by quantam effects in curved space time. The need for dark energy may be eradicated by this modified gravity and may aid for the fusion of the early time inflation and cosmic acceleration. Nojiri and Odinstov have studied about modified gravity and proposed some functional forms such as $ln(R)$ or $R^{-n}(lnR)^m$ and $R+\gamma R^{-n}(ln\frac{R}{\mu^2})^m$\cite{snoo,nodi}. Fayyaz and Shamir\cite{fay} have analysed wormhole structures in logarithmic-corrected $R^2$ gravity. Kourosh and Tahereh \cite{kou} have discussed phantom-like behavior in $f(R)= R + \beta log(\frac{R}{\mu^2})
+ \gamma R^m$ gravity.\\
\indent By appending the torsion scalar component to the exponential $f(R)$ theory\cite{cog,lin,ba,ya,che}, the functional form is $f(R,T)=R+\kappa e^{-\iota R}+\lambda T$ where $\kappa$, $\iota$ and $\lambda$ are constants. The reason behind choosing this functional form it comes up with the best way of exploring cosmic acceleration. In contrast to the $\Lambda CDM$  model the exponential gravity model has one more parameter included in it and it also permits the relaxation of fine tuning. Vinutha et al. \cite{kn} have studied on Bianchi type cosmological models in modified theory with exponential functional form.  Paul et al.\cite{pa} have worked on accelerating universe in modified theories of gravity. Sahoo et al.\cite{sah} have studied on $f(R,T)=f(R)+\lambda T$ gravity models as alternatives to cosmic acceleration. Moreas and Sahoo\cite{mos} have discussed traversable wormholes by using functional form $f(R,T) = R +\gamma e^{\chi T}$ and also with this functional form Moreas et al.\cite{mor} studied FRW cosmological model.\\
\indent When compared to other anisotropic metrics, Kantowski-Sachs model is very simple and easy to analyze. The cosmologies of Kantowski-Sachs metric possess two properties of symmetry such as spherical symmetry and invariance under spatial translation. It describes spatially homogeneous, anisotropic universe and interior of black holes that does not allow a simply transitive group of motions. It is also used to analyze the behavior of the added degrees of freedom in quantum cosmological model. This metric represents three different anisotropic $3+1$ dimensional space time and positive curvature models. The study of anisotropic models were nourished by the theoretical studies and observations of CMB which also been extended to modified theories of gravity. Thus this model with an anisotropic nature appeared most appropriate in describing the early stage of the cosmos. some of the authors who worked on Kantowski Sachs model are \cite{co,rena,vin,chau,wan,web,kat}.\\
\indent This article is organized as follows: In section 2, $f(R,T)$ gravity field equations are obtained and in section 3 the field equations of power-law, logarithmic and exponential functional forms are solved. Section 4 discusses the physical and geometrical properties of three cases using graphs and section 5 concludes our results.\\
\section{A brief review of $f(R,T)=f_1(R) + f_2(T)$ model}
The final action principle of $f(R, T)$ gravity which is a function of matter Lagrangian $L_m$ is read as
\begin{equation}
\label{g1}S= \int \left[\frac{1}{16\pi}f(R, T) +L_{m}\right] \sqrt{-g}d^{4}x,
\end{equation} \\
\hskip 0.6 cm where $g$ is the metric determinant of the fundamental tensor $g_{ij}$, $f(R,T)$ is an arbitrary function of  $R$ and $T$ which is mentioned in the abstract, $L_{m}$ is the usual matter Lagrangian density and we consider $G=c=1$.\\
By varying the above equation \eqref{g1} with respect to $g_{ij}$,we obtain the field equations of $f(R,T)$ gravity in covariant tensor form as
\begin{equation}
\label{g2} f_{R}(R,T)R_{ij}-\frac{1}{2}f(R,T)g_{ij} + (g_{ij} \Box -\nabla_{i}  \nabla_{j} )f_{R}(R,T) =8\pi T_{ij} - f_{T}(R,T)\theta_{ij}-f_{T}(R,T)T_{ij}, 
\end{equation}
here, $\nabla_{i}$ is the covariant derivative and $\Box=\nabla _i\nabla_ j$ is the D'Alemberts operator.
$f_R=\frac{\partial f(R,T)}{\partial R}$, $f_T=\frac{\partial f(R,T)}{\partial T}$ and $R_{ij}$ is the Ricci tensor,
where   
\begin{equation}
\label{g3}\theta_{ij}=-2T_{ij}+g_{ij}L_{m}-2g^{lk}\frac{\partial^2L_{m}}{\partial{g^{ij}g^{lk}}}.
\end{equation}
Here the energy-momentum tensor is considered to be a perfect fluid which is defined as 
\begin{equation}
\label{g5}T_{ij}=(p+\rho)u_iu_j-pg_{ij},
\end{equation} 
where $u_{i}$ denotes four velocity vector in co-moving coordinates i.e. $u_{i}=(1,0,0,0)$ and  $u_{i}u^j=1$. Hence, the components of energy-momentum tensor become $T_{ij}$=diag$(\rho, -p, -p, -p)$, where $p$ is the pressure and $\rho$ is the energy density of perfect fluid. Several authors have studied  by choosing energy-momentum tensor as perfect fluid which are included in the references\cite{srv,vas,ray,rao,rav,ravi,anp,anpp}. It takes the form by replacing  matter Lagrangian as $L_{m}=-p$\cite{be,bi,so} in equation \eqref{g3}.
\begin{equation}
\label{g4}\theta_{ij}=-2T_{ij}-pg_{ij}
\end{equation}
Consequently the field equations for $f(R,T)$ gravity are procured with the aid of $T=\rho-3p$ in equation \eqref{g2} as
\begin{equation}
\begin{aligned}
\label{g6}G_{ij}=&\frac{1}{ f_{R}(R,T)}\Big[[8\pi+f_{T}(R,T)]T_{ij}+pf_{T}(R,T)g_{ij}+\frac{1}{2}[f(R,T)-Rf_{R}(R,T)]g_{ij}\\&-(g_{ij}\Box-\nabla_{i} \nabla_j )f_{R}(R,T)\Big],
\end{aligned}
\end{equation}
where $G_{ij}$ is the Einstein tensor which is expressed as $R_{ij}-\frac{1}{2}Rg_{ij}$.\\ 
Here, we consider the functional form $f(R,T) = f_1(R) + f_2(T)$  i.e. 
\begin{equation}
\label{g8}f(R,T)=R+\gamma R^2-\frac{\mu^4}{R}+\lambda T
\end{equation}
\begin{equation}
\label{g9} f(R,T)=R+\nu ln(\tau R)+\lambda T
\end{equation}
\begin{equation}
\label{g10} f(R,T)=R+\kappa e^{-\iota R}+\lambda T
\end{equation}
as case I, II and III respectively.
\section{Metric and solutions of the field equations}
Now the metric takes the form,
\begin{equation}
\label{g11}ds^{2}=dt^{2}-M^{2}(t)dr^{2}- N^{2}(t)(d\theta^{2}+ \sin^{2}\theta d\psi^{2},
)
\end{equation}
where $M$ and $N$ are metric potentials and functions of cosmic time $t$ only and co-moving coordinates are $(r,\theta,\psi)$.\\ 
\subsection{Case I - (negative and positive powers of curvature)}
 The functional form $f(R,T)=R+\gamma R^2-\frac{\mu^4}{R}+\lambda T$ field equations are as follows
\begin{equation}
\hskip -2 cm\label{g16}\left.
\begin{aligned}
\frac{1}{N^2}+\frac{2\ddot{N}}{N}+\frac{\dot{N^2}}{N^{2}}&=-\frac{(8\pi+\frac{3\lambda}{2})p}{1+2R\gamma+\frac{\mu^4}{R^2}}+\frac{\lambda\rho}{2(1+2R\gamma+\frac{\mu^4}{R^2})}-\frac{(\frac{\gamma R^2}{2}+\frac{\mu^4}{R})}{1+2R\gamma+\frac{\mu^4}{R^2}}\\ &\hskip 2.6 cm- \frac{(2\gamma-\frac{2\mu^4}{R^3})}{1+2R\gamma+\frac{\mu^4}{R^2}}\Big[\frac{2\dot{N}}{N}\dot{R}+\ddot{R}\Big]-\frac{\frac{6\mu^4\dot{R}^2}{R^4}}{1+2R\gamma+\frac{\mu^4}{R^2}}.
\end{aligned}\right\}
\end{equation}
\begin{equation}
\hskip  -2 cm\label{g17}\left.
\begin{aligned}
\frac{\ddot{M}}{M}+\frac{\ddot{N}}{N}+\frac{\dot{M}\dot{N}}{{M}{N}}&=-\frac{(8\pi+\frac{3\lambda}{2})p}{1+2R\gamma+\frac{\mu^4}{R^2}}+\frac{\lambda\rho}{2(1+2R\gamma+\frac{\mu^4}{R^2})}-\frac{(\frac{\gamma R^2}{2}+\frac{\mu^4}{R})}{1+2R\gamma+\frac{\mu^4}{R^2}}\\ &\hskip 1.2 cm-\frac{(2\gamma-2\frac{\mu^4}{R^3})}{1+2R\gamma+\frac{\mu^4}{R^2}}\Big[(\frac{\dot{M}}{M}+\frac{\dot{N}}{N})\dot{R}+\ddot{R}\Big]-\frac{6\frac{\mu^4\dot{R}^2}{R^4}}{1+2R\gamma+\frac{\mu^4}{R^2}}.
\end{aligned}\right\}
\end{equation}
\begin{equation}
\hskip -2cm\label{g18} \left.
\begin{aligned}
2\frac{\dot{M}\dot{N}}{{M}{N}} +\frac{\dot{N}^{2}}{N^{2}}+ \frac{1}{N^{2}} &=\frac{(8\pi+\frac{3\lambda}{2})\rho}{1+2R\gamma+\frac{\mu^4}{R^2}}-\frac{\lambda p}{2(1+2R\gamma+\frac{\mu^4}{R^2})}-\frac{(\frac{\gamma R^2}{2}+\frac{\mu^4}{R})}{1+2R\gamma+\frac{\mu^4}{R^2}}\\& \hskip 4.5 cm-\frac{(2\gamma-2\frac{\mu^4}{R^3})}{1+2R\gamma+\frac{\mu^4}{R^2}}\Big[\frac{\dot{M}}{M}+\frac{2\dot{N}}{N}\Big]\dot{R}.
\end{aligned}\right\}
\end{equation}
here dot denotes derivate  with respect to $t$.\\
\subsection{Case - II (logarithmic curvature)}
Field equations corresponding to the $f(R,T)=R+\nu ln(\tau R)+\lambda T$ are
\begin{equation}
\hskip -2 cm\label{24}\left.
\begin{aligned}
\frac{1}{N^2}+\frac{2\ddot{N}}{N}+\frac{\dot{N^2}}{N^{2}}&=\frac{-(8\pi+\frac{3\lambda}{2})p}{1+\frac{\nu}{R}}+\frac{\lambda\rho}{2(1+\frac{\nu}{R})}-\frac{\nu(1-ln(\tau R))}{2(1+\frac{\nu}{R})}\\ &\hskip 3cm+\Big[\frac{2\dot{N}}{N}\dot{R}+\ddot{R}\Big]\frac{\frac{\nu}{R^2}}{1+\frac{\nu}{R}}-\frac{\frac{2\nu\dot{R}^2}{R^3}}{1+\frac{\nu}{R}}.
\end{aligned}\right\}
\end{equation}
\begin{equation}
\hskip -2 cm\label{25}\left.
\begin{aligned}
\frac{\ddot{M}}{M}+\frac{\ddot{N}}{N}+\frac{\dot{M}\dot{N}}{{M}{N}}&=-\frac{(8\pi+\frac{3\lambda}{2})p}{1+\frac{\nu}{R}}+\frac{\lambda\rho}{2(1+\frac{\nu}{R})}-\frac{\nu(1-ln(\tau R))}{2(1+\frac{\nu}{R})}\\ & \hskip 2 cm+\frac{\frac{\nu}{R^2}}{1+\frac{\nu}{R}}\Big[(\frac{\dot{M}}{M}+\frac{\dot{N}}{N})\dot{R}+\ddot{R}\Big]-\frac{\frac{2\nu\dot{R}^2}{R^3}}{1+\frac{\nu}{R}}.\\
\end{aligned}\right\}
\end{equation}
\begin{equation}
\hskip -2 cm\label{26}\left.
\begin{aligned}
2\frac{\dot{M}\dot{N}}{{M}{N}} +\frac{\dot{N}^{2}}{N^{2}}+ \frac{1}{N^{2}} =\frac{(8\pi+\frac{3\lambda}{2})\rho}{1+\frac{\nu}{R}}-\frac{\lambda p}{2(1+\frac{\nu}{R})}-\frac{\nu(1-ln(\tau R))}{2(1+\frac{\nu}{R})}\\ & \hskip -4.5 cm+\frac{\frac{\nu}{R^2}}{1+\frac{\nu}{R}}\Big[\frac{\dot{M}}{M}+\frac{2\dot{N}}{N}\Big]\dot{R}.
\end{aligned}\right\}
\end{equation}
\subsection{ Case - III (exponential curvature)}
Field equations corresponding to the $f(R,T)=R+\kappa e^{-\iota R}+\lambda T$ are given as follows:
\begin{equation}
\label{30}\left.
\begin{aligned}
\frac{1}{N^2}+\frac{2\ddot{N}}{N}+\frac{\dot{N^2}}{N^{2}}&=-\frac{(8\pi+\frac{3\lambda}{2})p}{1-\kappa \iota e^{-\iota R}}+\frac{\lambda\rho}{2(1-\kappa \iota e^{-\iota R})}+\frac{\kappa e^{-\iota R}(1+\iota R)}{2(1-\kappa \iota e^{-\iota R})}\\ &\hskip 1 cm-\frac{\kappa \iota^2e^{-\iota R}}{1-\kappa \iota e^{-\iota R}}\Big[(\frac{2\dot{N}}{N})\dot{R}+\ddot{R}\Big]+\frac{\kappa \iota^3e^{-\iota R}\dot{R}^2}{1-\kappa \iota e^{-\iota R}}.\\
\end{aligned}\right\}
\end{equation}
\begin{equation}
\label{31}\left.
\begin{aligned}
\frac{\ddot{M}}{M}+\frac{\ddot{N}}{N}+\frac{\dot{M}\dot{N}}{{M}{N}}&=-\frac{(8\pi+\frac{3\lambda}{2})p}{1-\kappa \iota e^{-\iota R}}+\frac{\lambda\rho}{2(1-\kappa \iota e^{-\iota R})}+\frac{\kappa e^{-\iota R}(1+\iota R)}{2(1-\kappa \iota e^{-\iota R})}\\ &\hskip 0.2cm-\frac{\kappa \iota^2e^{-\iota R}}{1-\kappa \iota e^{-\iota R}}\Big[(\frac{\dot{M}}{M}+\frac{\dot{N}}{N})\dot{R}+\ddot{R}\Big]+\frac{\kappa \iota^3e^{-\iota R}\dot{R}^2}{1-\kappa \iota e^{-\iota R}}.\\
\end{aligned}\right\}
\end{equation}
\begin{equation}
\label{32}\left.
\begin{aligned}
2\frac{\dot{M}\dot{N}}{{M}{N}} +\frac{\dot{N}^{2}}{N^{2}}+ \frac{1}{N^{2}} &=\frac{(8\pi+\frac{3\lambda}{2})\rho}{1-\kappa \iota e^{-\iota R}}-\frac{\lambda p}{2(1-\kappa \iota e^{-\iota R})}+\frac{\kappa e^{-\iota R}(1+\iota R)}{2(1-\kappa \iota e^{-\iota R})}\\ & \hskip 2 cm-\frac{\kappa \iota^2e^{-\iota R}\dot{R}}{1-\kappa \iota e^{-\iota R}}\Big[\frac{\dot{M}}{M}+\frac{2\dot{N}}{N}\Big].
\end{aligned}\right\}
\end{equation}
To obtain solutions for highly non-linear equations is very strenuous and in order to remove such complications we require some constraints.\\
(i) We consider $\sigma$ is proportional to  $\theta$(where $\sigma$ is the shear scalar and $\theta$ is the expansion scalar) and it generate linear relationship between two metric potentials in terms of $M$ and $N$ as
\begin{equation} 
\label{g54a} M = N^n,  
\end{equation}
$n\not= 0, 1$ is constant. The physical motivation for assuming this condition is that Hubble expansion of the universe may attain isotropy by the observations of the velocity redshift relation for extra galatic sources if the value of $\frac{\sigma}{\theta}$ is constant\cite{sac}.\\
(ii) The average scale factor is assumed as a hyperbolic expansion
\begin{equation}
\label{36}a(t) = \sinh(\alpha t)^\frac{1}{\beta},
\end{equation} 
where $\alpha>0$, ${\beta}>0 $  are constants. The consequence of using this scale factor is time dependent deceleration parameter $q$ \cite{chaw}. This average scale factor tends to zero if $t \to 0$ and if $t \to \infty$ then $a(t)$ becomes infinity.
\\The directional Hubble parameters are 
\begin{equation} 
\label{g54b} H_{1} = \frac{\dot M}{M}~~~~~~~~ H_{2}=  H_{3} = \frac{\dot N}{N}.
\end{equation}
The average Hubble parameter is,
\begin{equation}
\label{g54c}   H =\frac{1}{3}(H_{1}+2H_{2}). 
\end{equation}
By substituting equation \eqref{g54c} in equation \eqref{g54b}, we get
 \begin{equation}
 \label{g66}   H =\frac{\dot{a}}{a}=\frac{1}{3}\Big(\frac{\dot M}{M}+\frac{\dot N}{N}\Big).
 \end{equation}
From equations \eqref{g54a} - \eqref{g66}, we obtain metric potentials of $M$ and $N$ as 
\begin{equation}
\label{g67}
 M = (\sinh(\alpha t))^\frac{3n}{\beta(n+2)}, 
\end{equation}
\begin{equation}
\label{g68}
  N = (\sinh(\alpha t))^\frac{3}{\beta(n+2)}. \\
\end{equation}
If $t \to \infty$ then $M$ and $N$ are nonzero, hence, our model is free from singularity.\\ 
Using equations \eqref{g67} and \eqref{g68}, the Kantowski sachs metric obtained as 
\begin{equation}
\label{g54d}ds^{2}=dt^{2}-(\sinh(\alpha t))^\frac{6n}{\beta(n+2)}\; dr^{2}- (\sinh(\alpha t))^\frac{6}{\beta(n+2)}(d\theta^{2}+ \sin^{2}\theta d\psi^{2}).
\end{equation}
The above metric represents a perfect fluid Kantowski-Sachs universe in $f(R, T)$ theory of gravity.\\
\subsection{Pressure and energy density for case I}
By solving the equations of \eqref{g16},\eqref{g17} and \eqref{g18} we get the pressure of the model as
\begin{equation}
\begin{aligned}         
\label{g19} p={\frac{1}{4}}\Big({{\frac{\chi+{\xi}-{2}{\eta}-{\phi_4}+{2}{\phi_5}+{2}{\phi_6}-{\phi_7}}{\phi_2-\phi_1}}-{\frac{{\chi}+{\xi}+{2\eta}+{4}{\phi_3}+{7}{\phi_4}+{2}{\phi_5}+{2}{\phi_6}+{3}{\phi_7}}{\phi_1+\phi_2}}}\Big), 
\end{aligned}        
\end{equation} 
and the energy density of the model is obtained as
\begin{equation}
\begin{aligned}         
\label{g20} \rho={\frac{1}{4}}\Big({\frac{{\chi}+{\xi}+{2\eta}+{4}{\phi_3}+{7}{\phi_4}+{2}{\phi_5}+{2}{\phi_6}+{3}{\phi_7}}{\phi_1+\phi_2}}+{{\frac{\chi+\xi-{2}{\eta}-{\phi_4}+{2}{\phi_5}+{2}{\phi_6}-{\phi_7}}{\phi_2-\phi_1}}}\Big), 
\end{aligned}        
\end{equation}
\subsection{Pressure and energy density for case II}
By solving the equations \eqref{24},\eqref{25} and \eqref{26} we get the expression for pressure is
\begin{equation}
\begin{aligned}         
\label{g27} p={\frac{1}{4}}\Big({{\frac{\chi+\xi-{2}{\eta}+{\phi_4}-{2}{\phi_5}+{4}{\phi_6}+{\phi_7}}{\phi_2-\phi_1}}-{\frac{{\chi}+{\xi}+{2\eta}+{4}{\phi_3}-{7}{\phi_4}-{2}{\phi_5}+{4}{\phi_6}-{3}{\phi_7}}{\phi_1+\phi_2}}}\Big), 
\end{aligned}        
\end{equation} 
and the energy density of the model is obtained as
\begin{equation}
\begin{aligned}         
\label{g28} \rho={\frac{1}{4}}\Big({\frac{{\chi}+{\xi}+{2\eta}+{4}{\phi_3}-{7}{\phi_4}-{2}{\phi_5}+{4}{\phi_6}-{3}{\phi_7}}{\phi_1+\phi_2}}+{{\frac{\chi+\xi-{2}{\eta}+{\phi_4}-{2}{\phi_5}+{4}{\phi_6}+{\phi_7}}{\phi_2-\phi_1}}}\Big), 
\end{aligned}        
\end{equation}
\subsection{Pressure and energy density for case III} 
By solving the equations \eqref{30}, \eqref{31} and\eqref{32} we get the pressure of the model as
\begin{equation}
\begin{aligned}         
\label{g33} p={\frac{1}{4}}\Big({{\frac{\chi+\xi-{2}{\eta}-{\phi_4}+{2}{\phi_5}-{2}{\phi_6}-{\phi_7}}{\phi_2-\phi_1}}-{\frac{{\chi}+{\xi}+{2\eta}-{4}{\phi_3}+{7}{\phi_4}+{2}{\phi_5}-{2}{\phi_6}+{3}{\phi_7}}{\phi_1+\phi_2}}}\Big), 
\end{aligned}        
\end{equation} 
and the energy density of the model is obtained as
\begin{equation}
\begin{aligned}         
\label{g34} \rho={\frac{1}{4}}\Big({\frac{{\chi}+{\xi}+{2\eta}-{4}{\phi_3}+{7}{\phi_4}+{2}{\phi_5}-{2}{\phi_6}+{3}{\phi_7}}{\phi_1+\phi_2}}+{{\frac{\chi+\xi-{2}{\eta}-{\phi_4}+{2}{\phi_5}-{2}{\phi_6}-{\phi_7}}{\phi_2-\phi_1}}}\Big), 
\end{aligned}        
\end{equation}
The values of $\chi$, $\xi$ and $\eta$ are same for all the three cases whereas, the values of  $\phi_i$, for i=1 to 7 for the corresponding cases are clearly given in appendix section.
\section{Physical and geometrical properties}
The average Hubble parameter is 
\begin{equation}\label{69}
 H=\frac{\alpha}{\beta}\coth(\alpha t).
\end{equation} 
From the figure of Hubble parameter, we trace that it decreases with the decrease of redshift i.e. decreases as time increases. By choosing the values of $\alpha =0.21$ and $\beta =3.10$ in the scale factor the Hubble parameter is obtained as 0.07$Gyr{s^{-1}}$ which is nearly equal to the present observational data \cite{god}. For this quantity the dimension is $\frac{1}{time}$. By using this formula, we can also measure the age of the cosmos.\\
\begin{figure}[h]
	\centering
	\includegraphics[width=0.5\linewidth]{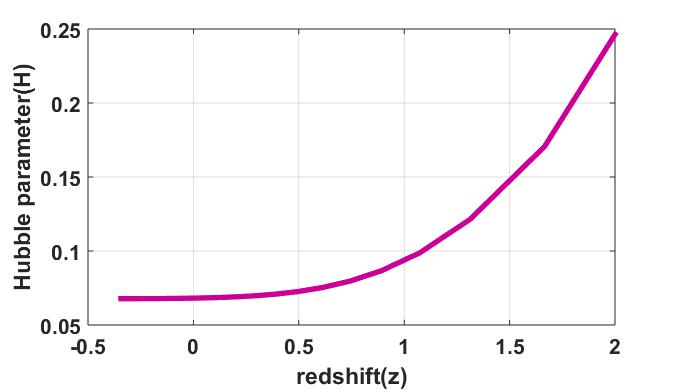}
	\caption{Plot of Hubble parameter($H$) versus redshift($z$)}
	\label{fig:u}
\end{figure}
(ii) The volume of the model is given by
\begin{equation}\label{63}
	V = a^3 = (\sinh(\alpha t))^\frac{3}{\beta}.
\end{equation}
In figure 2, it is clear that the spatial volume increases with the decrease of redshift i.e. it increases as the time increases and is finite at final epoch.\\
\begin{figure}[h]
	\centering
	\includegraphics[width=0.5\linewidth]{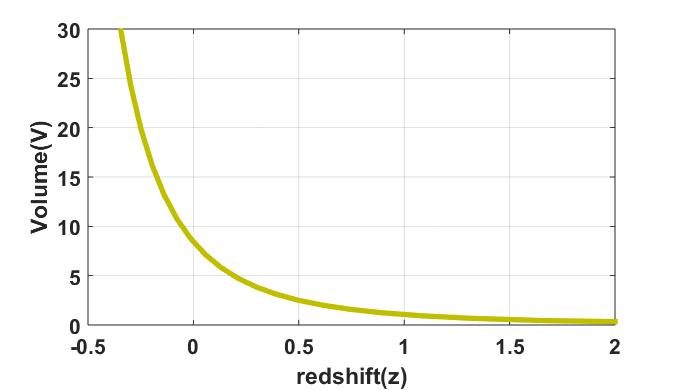}
	\caption{Plot of volume($V$) versus redshift($z$)}
	\label{fig:v}
\end{figure}
(iii) The expansion scalar $\theta$ is
\begin{equation}   
\label{g54e}  \theta =u^{i}_{;i}= 3H= \frac{3\alpha \coth(\alpha t)}{\beta}.
\end{equation}
From figure 3, it is observed that expansion scalar  decreases with the decrease of redshift i.e. it decreases as time increases. Here we noticed that for $t=0$ the expansion scalar is infinite.\\
\begin{figure}[h]
	\centering
	\includegraphics[width=0.5\linewidth]{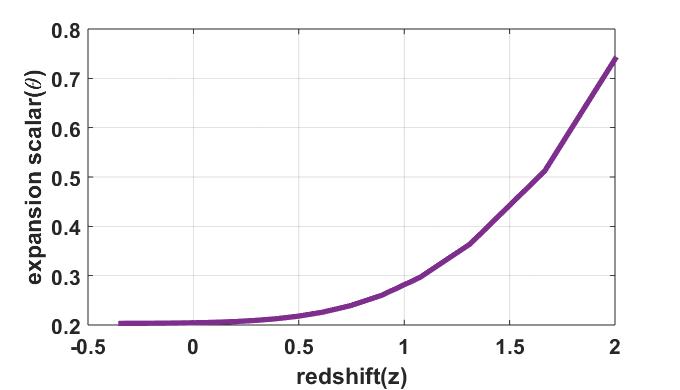}
	\caption{Plot of expansion scalar($\theta$) versus redshift($z$)}
	\label{fig:e}
\end{figure}
(iv) We get the shear scalar as  
\begin{equation}
\begin{aligned}         
\label{g54f}\sigma^{2}=\frac{3\alpha^2(n-1)^2 \coth^2(\alpha t)}{\beta^2(n+2)^2}, 
\end{aligned}        
\end{equation}
when $t=0$, $\sigma^{2}$ (shear scalar) tends to infinity.\\
(v) The mean anisotropy parameter $A_h$ is obtained as 
\begin{equation} 
\label{62}  A_h=\frac{1}{3}\Big[\sum_{i=1}^{3}\Big(\frac{H_i-H}{H}\Big)^2\Big]
\end{equation}
where $i=1, 2, 3$ indicate the directional Hubble parameters for the coordinates of $r, \theta$ and $\psi$. The mean anisotropy parameter is defined on the basis of directional Hubble parameter and mean Hubble parameter. 
\begin{equation}
\label{g54g} A_h=\frac{2(n-1)^2}{(n+2)^2};  n \not= -2.
\end{equation} 
The mean anisotropy parameter $A_h$ is useful for checking if the model is anisotropic or not. In the present model $A_h=0$ for $n=1$ and $A_h \not= 0$ for $n \not= 1$ that is the model is anisotropic for $n \not= 1$ and isotropic for $n=1$.\\
In all the discussions and graphical representation of physical parameters we constraint the constants for case I as $\alpha=0.21$, $\beta=3.10$, $n=7.38$, $\lambda=-10.02$, $\mu=0.2$, $\gamma=0.03$, case II as  $\nu=0.001$, $\tau=0.002$ and case III as $\kappa=0.2$, $\iota=0.009$. The values of parameters $\alpha$, $\beta$, $n$, $\lambda$ in cases II and III are same as that of Case I.\\
(iv) The deceleration parameter is 
\begin{equation}
\label{g54h}q =-1+\frac{d}{dt}(\frac{1}{H}). 
\end{equation}
In this model by using hyperbolic function we obtained deceleration parameter as
\begin{equation}
\begin{aligned}
\label{g54i}q = -1+\beta(1-\tanh^2(\alpha t)).
\end{aligned}
\end{equation}
    When $t < \frac{1}{\alpha}\tanh^{-1}(1-\frac{1}{\beta})^{\frac{1}{2}}$, $q$ has negative value which represents that the universe is accelerating whereas if $t > \frac{1}{\alpha} \tanh^{-1}(1- \frac{1}{\beta})^{\frac{1}{2}}$, $q$ has positive value which represents that the universe is decelerating. The quantities such as  $q$ and $H$ specifies the geometric properties of the cosmos. \\
v)Throughout the plots uniform colouring is followed by giving the colours brown for pressure, navy blue for energy density, sky blue for EoS parameter,   blue for SEC, green for NEC and red for DEC. Figures 4,5 and 6 illustrate the variation of pressure against redshift  in cases I, II and III respectively. The figures shows that in three cases pressure is negative and it is known that a negative pressure fluid is the correct mechanism which is capable of explaining cosmic acceleration within the standard cosmologies, despite the fact that in the latter it is necessary to bring the cosmological constant to get this exotic characteristic. In pressure graphs increase with the decrease of redshift that it is increases as the time increases is perceived which represents cosmic acceleration.\\
\begin{figure}[ht!]
\begin{minipage}{0.32\textwidth}
\includegraphics[width=1.1\linewidth]{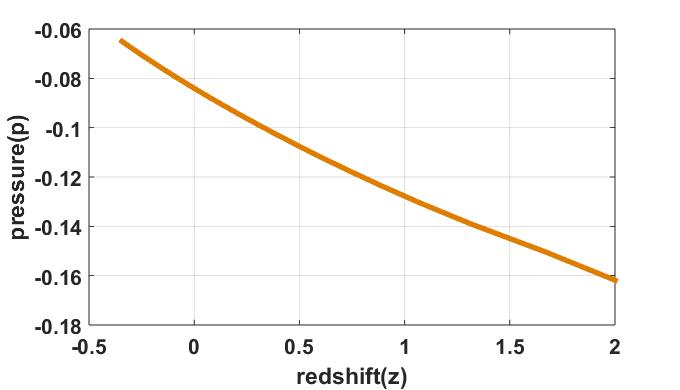}
\caption{Pressure($p$) \\in case I}
\label{fig:p1}
\end{minipage}
\begin{minipage}{0.32\textwidth}
	\centering
	\includegraphics[width=1.1\linewidth]{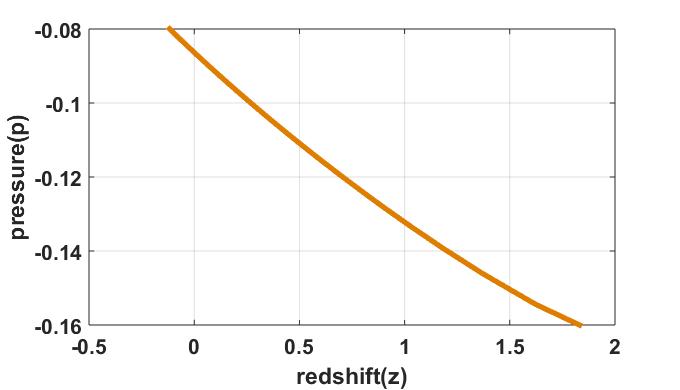}
	\caption{Pressure($p$) \\in case II}
	\label{fig:p2}
\end{minipage}
\begin{minipage}{0.32\textwidth}
	\centering
	\includegraphics[width=1.1\linewidth]{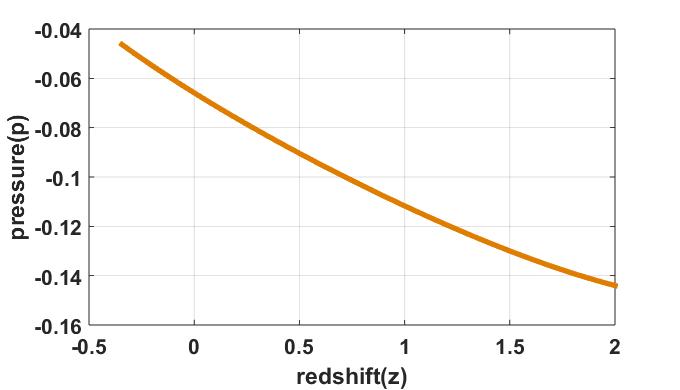}
	\caption{Pressure($p$) \\in case III}
	\label{fig:p3}
\end{minipage}
\end{figure}\\
vi) Figures 7,8 and 9 shows the evolution of energy density for cases I, II and III respectively. In all the cases the density  decreases with the decrease of redshift i.e. decreases as time increases.\\
\begin{figure}[ht!]
\begin{minipage}{0.32\textwidth}
	\centering	
	\includegraphics[width=1.1\linewidth]{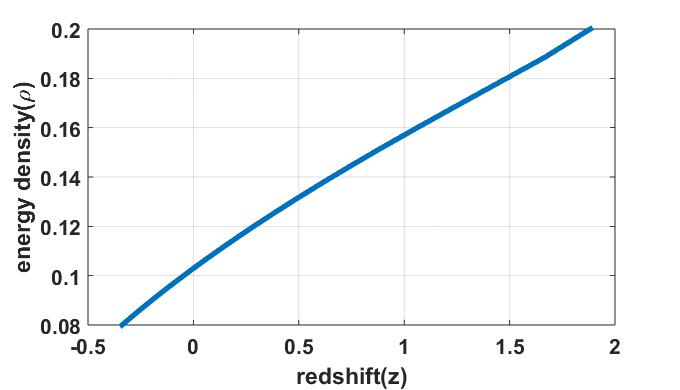}
	\caption{Energy density($\rho$)\\ in case I}
	\label{fig:rho2}
\end{minipage}
\begin{minipage}{0.32\textwidth}
	\centering
	\includegraphics[width=1.1\linewidth]{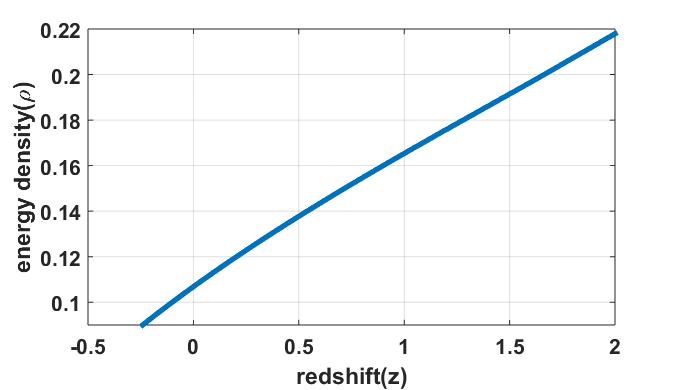}
	\caption{Energy density($\rho$)\\ in case II}
	\label{fig:rho1}
	\end{minipage}
\begin{minipage}{0.32\textwidth}
	\centering
	\includegraphics[width=1.1\linewidth]{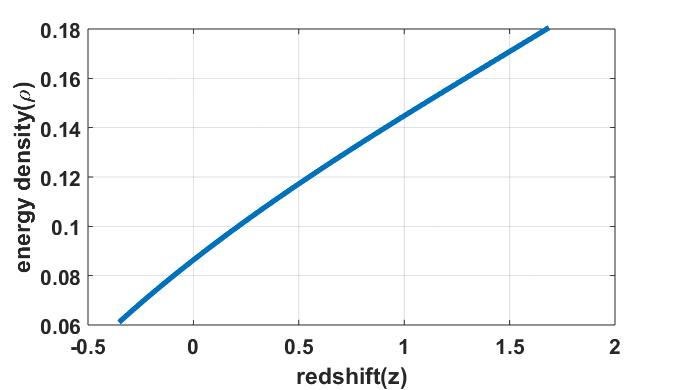}
	\caption{Energy density($\rho$)\\ in case III}
	\label{fig:rho3}
\end{minipage}
\end{figure}
vii) With great efforts the equation of state(EoS) parameter in cosmology of different dark energy models are examined.  The parameter relating to the equation of state is a dimensionless term that represents the matter state under some particular physical grounds. In the terminology of $p$ and $\rho$ the EoS can be interpret in the from of $\omega=\frac{p}{\rho}$. The EoS parameter  is distinguished in three regions namely quintessence, phantom, and quintom according to its range. In quintessence region the EoS paramter lies in the range of $-1<\omega<-\frac{1}{3}$, in phantom phase the EoS parameter is in the range of less than -1 (i.e. $\omega<-1$) and  in quintom $\omega=-1$. Figures 10, 11 and 12  of EoS parameter are drawn against redshift and observe that decrease with the decrease of redshift that is decrease as time increases. From the graphs we noticed that our model lies in quintessence region in three cases. According to Planck+nine years WMAP the current value of EoS parameter is approximately as $\omega=-1.13^{+0.24}_{-0.25}$\cite{ade}, and from SNe Ia data with galaxy clustering, CMBR anisotropy statistics the EoS parameter lies in the range $-1.33<\omega<-0.79$, $-1.67<\omega<-0.62$\cite{teg} respectively. From the figures of EoS parameters, it is seen that  three cases are approximately coincide with observational data which is a good result. \\
\begin{figure}[h]
	\begin{minipage}{0.32\textwidth}
	\centering
		\includegraphics[width=1.1\linewidth]{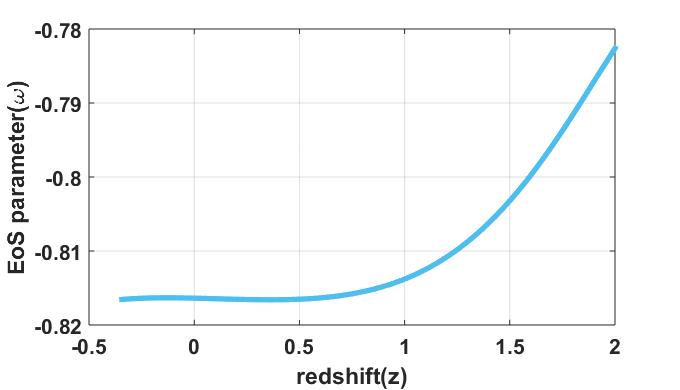}
	\caption{EoS parameter($\omega$) \\ in case I}
	\label{fig:w2}
\end{minipage}
\begin{minipage}{0.32\textwidth}
	\centering
	\includegraphics[width=1.1\linewidth]{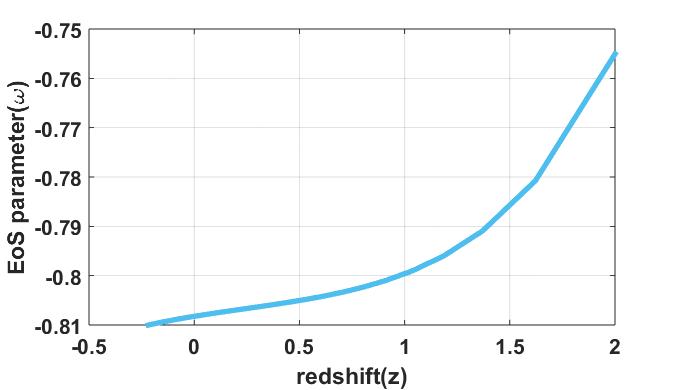}
	\caption{EoS parameter($\omega$)\\ in case II}
	\label{fig:wl}
\end{minipage}
\begin{minipage}{0.32\textwidth}
	\centering
	\includegraphics[width=1.1\linewidth]{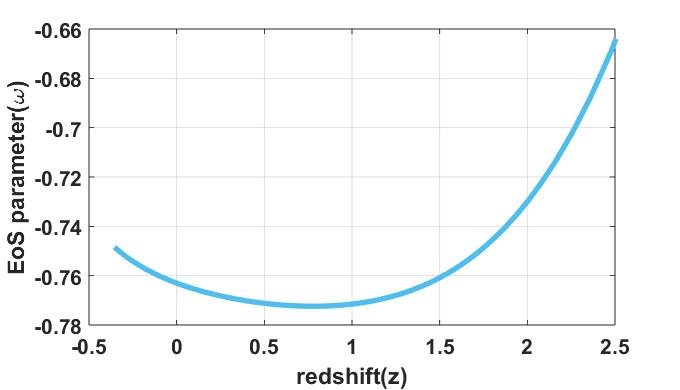}
	\caption{EoS parameter($\omega$)\\ in case III}
	\label{fig:w3}
	\end{minipage}
\end{figure}
viii)In modified theories of gravity, energy conditions\cite{cap,capo,lobo} plays a crucial role in studying the behaviour of spacelike and timelike geodesics and these conditions are came from Raychaudhuri equations\cite{rayc}. Energy conditions can be defined in many ways, such as geometric way and physical way. Moreover energy conditions are significant in the black hole physics, as they lay foundations of the singularity theorems. Another advantage of energy condition is that it allows basic tools to consider certain ideas about black holes and wormholes. There are four most commonly used fundamental energy conditions. The general expressions for energy conditions in regard of pressure and energy density are  given below:\\
(i)SEC(Strong Energy condition): Gravity always has to be attractive, and in cosmology $\rho + 3p \geq 0$, $\rho + p \geq 0$ should be observed.\\
(ii)DEC(Dominant Energy Condition): The energy density should always be positive when measured by any observer that is $\rho \geq 0$, $\rho \pm p \geq 0$, must be obeyed.\\
(iii)WEC(Weak Energy Condition): The energy density must always be positive when measured by any observer that is $\rho \geq 0$, $\rho + p \geq 0$.\\
(iv)NEC(Null Energy Condition): NEC is expressed in the form of $\rho +p \geq 0$ and it ensures the validity of second law of black hole thermodynamics.\\
Where NEC, WEC, DEC and SEC represents null, weak, dominant and strong energy conditions.
 According to present cosmological data to represent the universe with cosmic expansion the SEC of that model should be violated ($\rho+3p\leq 0$). For the obtained models the same scenario can be clearly observed from figures 13 to 15.  When compared to strong energy condition null energy condition is more beneficial, as it can be used algebraically due to its weakest pointwise energy condition which results in the strongest theorems and all these energy conditions, are met by electromagnetic field. From figures 16 to 18 it is clear that NEC($\rho+p \geq 0$) is satisfied in all the three cases for the obtained model. If NEC satisfies then the parameter EoS occurs in quintessence region. Also from figures 19 to 21 it is clear that DEC ($\rho+p \geq 0$) is fulfilled in all the three cases for the obtained model.
\begin{figure}[h]
	\begin{minipage}{0.32\textwidth}
	\centering
	\includegraphics[width=1.1\linewidth]{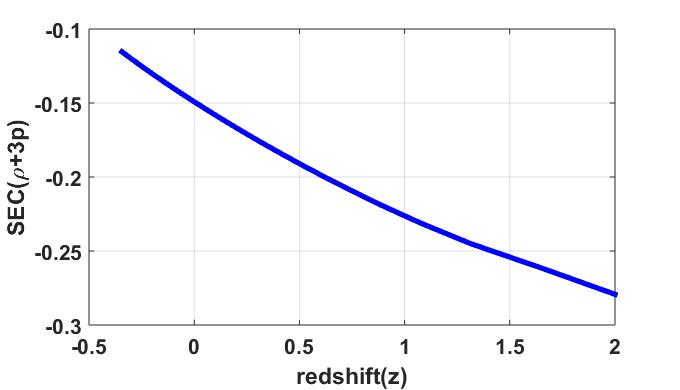}
	\caption{SEC in case I}
	\label{fig:ec}
\end{minipage}
\begin{minipage}{0.32\textwidth}
	\centering
	\includegraphics[width=1.1\linewidth]{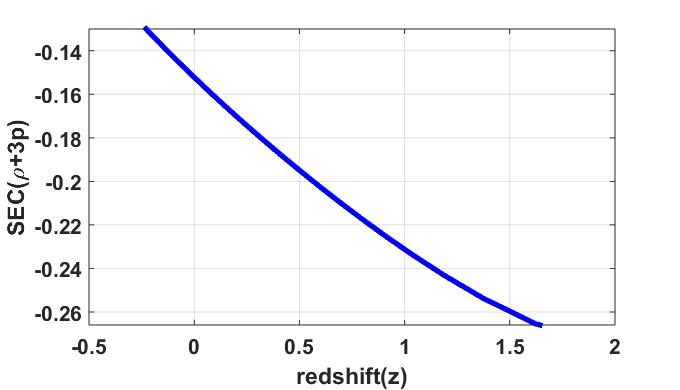}
	\caption{SEC in case II}
	\label{fig:ec1}
	\end{minipage}
\begin{minipage}{0.32\textwidth}
	\centering
	\includegraphics[width=1.1\linewidth]{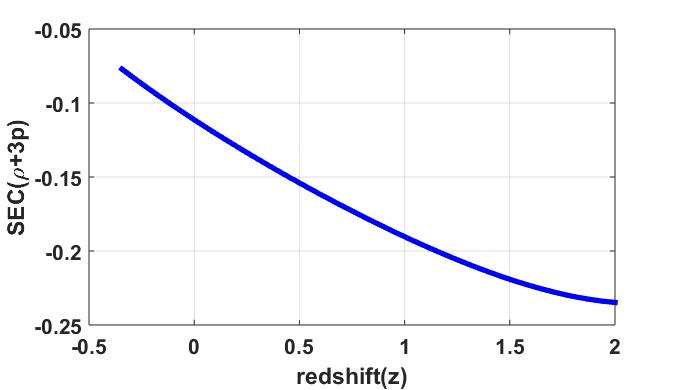}
	\caption{SEC in case III}
	\label{fig:ec2}
	\end{minipage}
\end{figure}
\begin{figure}[h]
	\begin{minipage}{0.32\textwidth}
		\centering
		\includegraphics[width=1.1\linewidth]{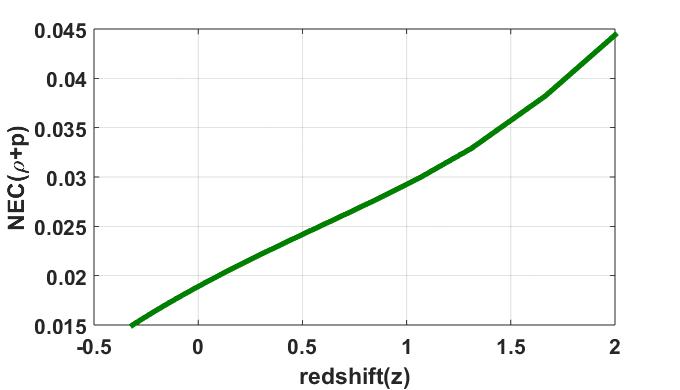}
		\caption{NEC in case I}
		\label{fig:ec21}
	\end{minipage}
	\begin{minipage}{0.32\textwidth}
		\centering
		\includegraphics[width=1.1\linewidth]{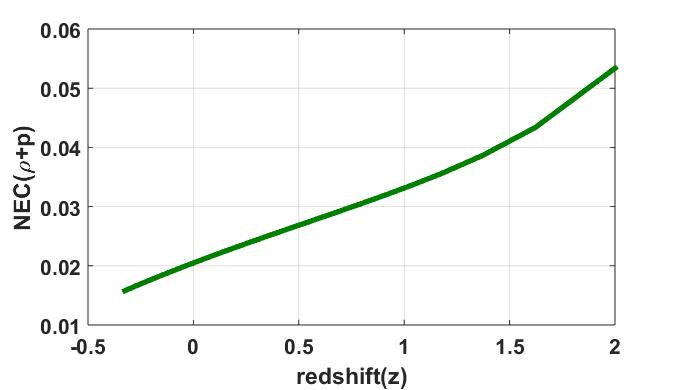}
		\caption{NEC in case II}
		\label{fig:ec12}
	\end{minipage}
	\begin{minipage}{0.32\textwidth}
		\centering
		\includegraphics[width=1.1\linewidth]{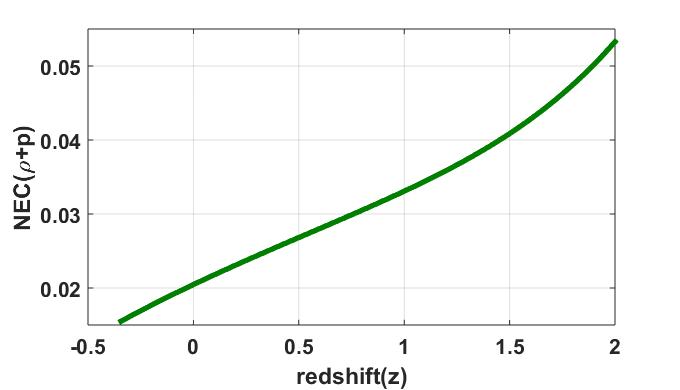}
		\caption{NEC in case III}
		\label{fig:ec32}
	\end{minipage}
\end{figure}
\begin{figure}[h]
	\begin{minipage}{0.32\textwidth}
		\centering
		\includegraphics[width=1.1\linewidth]{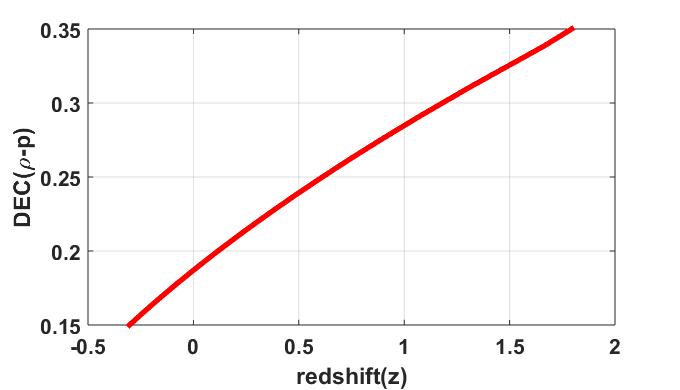}
		\caption{DEC in case I}
		\label{fig:ec13}
	\end{minipage}
	\begin{minipage}{0.32\textwidth}
		\centering
		\includegraphics[width=1.1\linewidth]{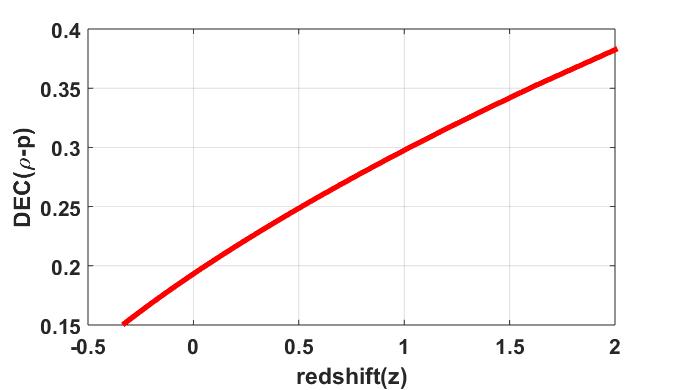}
		\caption{DEC in case II}
		\label{fig:ec23}
	\end{minipage}
	\begin{minipage}{0.32\textwidth}
		\centering
		\includegraphics[width=1.1\linewidth]{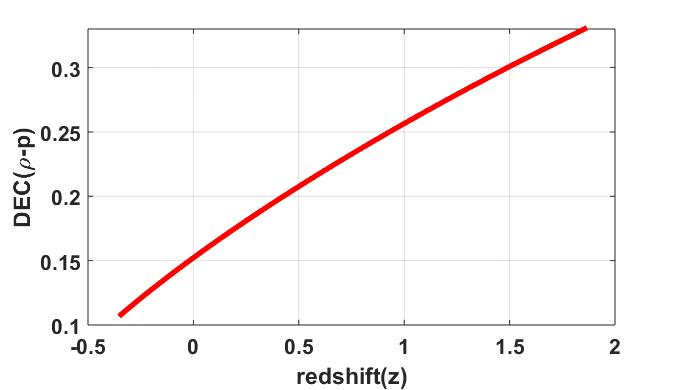}
		\caption{DEC in case III}
		\label{fig:ec33}
	\end{minipage}
\end{figure}
\subsection{Stability analysis}
Perturbations are essential for simplify a complex mathematical problems. There are several types of perturbations such as isotropic, anisotropic, homogeneous/inhomogeneous scalar, vector and tensor perturbations. The technique of perturbation is studied as a tool for finding approximate solution and comparing it to the obtained exact solution. Some of the researchers who studied on stability analysis are \cite{an,chi,bij}. Here the stability of solutions in terms of metric perturbation as following
\begin{equation}
\label{41}a_i \to a_{Bi}+\delta a_i=a_{Bi}(1+\delta b_i).
\end{equation}
The perturbation of volume scale factor, directional Hubble factors and mean Hubble factors are
\begin{equation}
\label{42}V \to V_B+V_B\sum_i \delta b_i,\\ \hskip 0.3cm \theta_i \to \theta_{Bi}+\sum_i \delta b_i,\\ \hskip 0.3cm \theta \to \theta_B+\frac{1}{3}\sum_i \delta b_i.
\end{equation}
The following equations are satisfied by the metric perturbation $\delta b_i$
\begin{equation}
\label{43}\sum_i  \delta \ddot b_i+2\sum_i \theta_{Bi}\delta \dot b_i=0,
\end{equation}
\begin{equation}
\label{44}\delta \ddot b_i+\frac{\dot V_B}{V_B}\delta \dot b_i+\sum_j \delta \dot b_j \theta_{Bi}=0,
\end{equation}
\begin{equation}
\label{45}\sum_i \delta \dot b_i=0.
\end{equation}
From equations \eqref{43} - \eqref{45}, we attain
\begin{equation}
\label{46}\delta \ddot b_i+\frac{\dot V_B}{V_B}\delta \dot b_i=0,
\end{equation}
where $V_B$ is the background spatial volume and for our case $V_B$ is
\begin{equation}
\label{47}V_B=(\sinh(\alpha t))^\frac{3}{\beta}.
\end{equation}
From above two equations, $\delta b_i$ is procured as 
  \begin{equation}
 \label{48}\delta b_i=c_1-c\Big(\frac{\beta \sqrt{cosh^2(\alpha t)} \sech(\alpha t) \sinh^{\frac{\beta-3}{\beta}}(\alpha t) {}_2 F_1{\left(\frac{1}{2}, \frac{\beta-3}{2 \beta}; \frac{3(\beta-1)}{2 \beta};-\sinh^2(\alpha t)\right)}}{\alpha(\beta-3)}\Big),
  \end{equation}
   where $c_1$ and $c$  are integrating constants.\\
   consequently, the actual fluctuations $\delta a_i=a_{Bi}\delta b_i$ is
   \begin{equation}
   \label{49}\delta a_i=\Big[c_1-c\Big(\frac{\beta \sqrt{cosh^2(\alpha t)} \sech(\alpha t) \sinh^{\frac{\beta-3}{\beta}}(\alpha t) {}_2 F_1{\left(\frac{1}{2}, \frac{\beta-3}{2 \beta}; \frac{3(\beta-1)}{2 \beta};-\sinh^2(\alpha t)\right)}}{\alpha(\beta-3)}\Big)\Big](\sinh(\alpha t))^\frac{-3}{\beta}.
   \end{equation}
   \begin{figure}[h]
   	\centering
   	\includegraphics[width=0.5\linewidth]{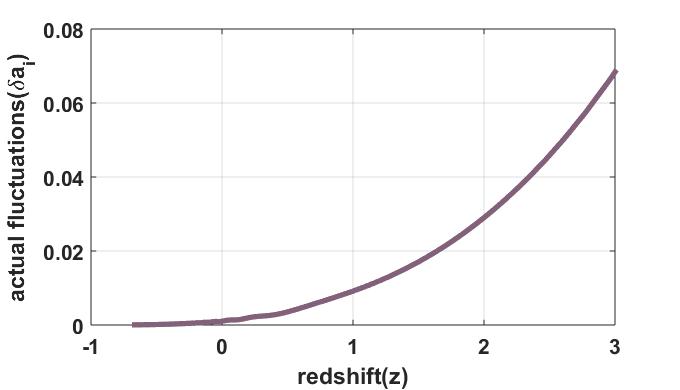}
   	\caption{Plot of actual fluctuations($\delta a_i$) versus redshift($z$)}
   	\label{fig:a}
   \end{figure}
Figure 22 shows the behaviour of  actual fluctuations versus redshift and it is noticed that it is a decreasing function with the decrease of redshift that is decreases as time increases. It is clear that $\delta a_i \to 0$ as $z \to -\infty$ and hence the background solution is shown to be stable against perturbation of gravitational field.
\section {Conclusions}
\hskip 0.6cm A cosmological model in $f(R,T)$ theory with three cases  namely power law, logarithmic and exponential curvature is obtained.  Hyperbolic scale factor is used to solve the field equations to get the solution in each case. The solutions of these field equations represent accelerating model of the universe. The graph of all parameters are drawn against redshift. In graphs the negative region of $z$ represents future epoch, positive region of $z$ represents past and $z=0$ indicates present. Obtained models are anisotropic and free from singularity all the way through the universe's evolution. By analyzing  all the parameters the conclusions are as follows:
\begin{itemize}
\item From figures 1 and 3 and from the equations \eqref{69} and \eqref{g54e} it can be seen that Hubble parameter and expansion scalar decreases with the decrease of redshift, and also it is clear that the Hubble parameter and expansion scalar are close to zero when $t \to \infty$.
\item From figure 2 it is clear that  volume  increases with the decrease of redshift which indicates volume of the expanding universe.  From equation \eqref{g54f}, it is noticed that the shear scalar is a function of time and tends to zero when $t \to \infty$. 
\item From equation \eqref{g54g}, the anisotropic parameter is independent of time and $A_h \not= 0$ for $n\not=1$, $A_h=0$ for $n=1$. But in this paper due to power law $n$ is different from one. Hence the models are anisotropic throughout.
\item From the graphs of pressure and energy density of all the three cases, it is clear  that the pressure and energy density are negative and positive respectively. Due to the negative pressure and positive energy density the universe is going through accelerating expansion. 
\item  The behavior of EoS parameter against redshift is represented in plots 11 to 13. From these graphs it is obvious that the model is in the quintessence region in three cases that is $-1<\omega < -\frac{1}{3}$ which  matches with present observational data.
\item In three cases, SEC is violated whereas NEC and DEC are fufilled. The violation of SEC leads to cosmic acceleration which is in good agreement with the  expansion of the cosmos.
\item As seen in the graph of stability analysis, the actual fluctuations begin with a small positive value and decreases to zero. As a result, the background solution is stable when the gravitational field is perturbed. 
\end{itemize}
A detailed discussion is provided through the obtained models for describing cosmic acceleration. Finally, through the detailed study of the models in three cases namely power law curvature $f(R,T)=R+\gamma R^2-\frac{\mu^4}{R}+\lambda T$, logarithmic curvature$ f(R,T)=R+\nu ln(\tau R)+\lambda T$ and exponential curvature $f(R,T)=R+\kappa e^{-\iota R}+\lambda T$ very good results which represents the universe accelerating expansion are observed. Moreover all the parameters discussed here matches with the recent observational data. At last, without existence of any exotic fluid, the current universe is accelerating is perceived in this paper which is a great outcome. As a future work, this work can be extended to other anisotropic models and can study the similarities and differences between them.\\
\section{Appendix}
The values of $\chi$, $\xi$, $\eta$ are same for all the three cases and are given below
$$
\begin{aligned}
\label{21}\hskip -8cm
\chi=\frac{\varrho_1}{\varrho_2}
\end{aligned}
$$
where 
$$
\begin{aligned}
\label{91a}\hskip -3cm
\varrho_{1}=\beta^2 \sinh(\alpha t)^2(\cosh(\alpha t)-1)(\cosh(\alpha t)+1)(n+2)^2\sinh(\alpha t)^\frac{-6+(-2n-4)\beta}{\beta(n+2)}\\-6(\frac{-9 \cosh(\alpha t)^2}{2}+\beta(n+2))\alpha^2
\end{aligned}
$$
$$
\begin{aligned}
\label{91b}\hskip -8cm
\varrho_{2}=\beta^2(n+2)^2\sinh(\alpha t)^2
\end{aligned}
$$
$$
\begin{aligned}
\label{22}
\hskip -8cm\xi=\frac{\varrho_3}{\varrho_2}
\end{aligned}
$$
where
$$
\begin{aligned}
\label{91c}\hskip -5.8cm
\varrho_3=-3\Big((-3n^2-3n-3)\cosh(\alpha t)^2+\beta(n+2)(n+1)\Big)\alpha^2
\end{aligned}
$$
$$
\begin{aligned}
\label{23}\hskip -8cm
\eta=\frac{\varrho_4}{\varrho_2}
\end{aligned}
$$
$$
\begin{aligned}
\label{91d}\hskip -3cm
\varrho_4=\beta^2 \sinh(\alpha t)^2(\cosh(\alpha t)-1)(\cosh(\alpha t)+1)(n+2)^2\sinh(\alpha t)^\frac{-6+(-2n-4)\beta}{\beta(n+2)}\\+18\alpha^2(n+\frac{1}{2})\cosh(\alpha t)^2
\end{aligned}
$$
\textbf{For case(i)} The values of $\phi_1$, $\phi_2$, $\phi_3$, $\phi_4$, $\phi_5$, $\phi_6$ and $\phi_7$ are given below \\
$$
\begin{aligned}
\label{g54}\hskip -8cm
\phi_1(t)=\frac{\varrho_5}{\varrho_6}
\end{aligned}
$$
where
$$
\begin{aligned}
\label{g56}\hskip 0.6cm
\varrho_5(t)=288(n+2)^2\sinh(\alpha t)^\frac{6+(2n+4)\beta}{\beta(n+2)}\beta^2\Big(-\Big((n+2)^2\beta-3n^2-6n-9\Big)\alpha^2\cosh(\alpha t)^2\sinh(\alpha t)^\frac{6}{\beta(n+2)}\\+\Big(\alpha^2\sinh(\alpha t)^\frac{6+(2n+4)\beta}{\beta(n+2)}+\frac{\cosh(\alpha t)^2\beta}{3}-\frac{\beta}{3}\Big)(n+2)^2\beta\Big)^2(\pi+\frac{3\lambda}{16})
\end{aligned}
$$
$$
\label{g57}
\begin{aligned}\hskip -4.8cm\varrho_6(t)&= \varrho_7(t)+4(\cosh(\alpha t)-1)(n+2)^2\beta^2(\cosh(\alpha t)+1)\varrho_8(t)
\end{aligned}
$$
$$
\label{91e}\left.
\begin{aligned}\hskip 0.6cm
\varrho_7(t)=\Big(\Big(\mu^4(n+2)^6\beta^6+324\alpha^4(n+2)^2(n^2+2n+3)^2\beta^2-11664\alpha^6(n^2+2n+3)^3\gamma\Big)\\& \hskip -13.6cm\cosh(\alpha t)^6-3(n+2)^2\beta\Big(\mu^4(n+2)^4\beta^5+72\alpha^4(n^2+2n+3)(n+2)^2\beta^2+108\alpha^4\\& \hskip -13.6cm(n^2+2n+3)^2\beta-3888\alpha^6(n^2+2n+3)^2\gamma\Big)\cosh(\alpha t)^4+3(n+2)^4\beta^2\Big(\mu^4(n+2)^2\beta^4\\& \hskip -13.6cm+12\alpha^4(n+2)^2\beta^2+72\alpha^4(n^2+2n+3)\beta-1296\alpha^6(n^2+2n+3)\gamma\Big)\cosh(\alpha t)^2\\& \hskip -13cm-\beta^3(n+2)^6(-432\alpha^6\gamma+\mu^4\beta^3+36\alpha^4\beta)\Big)\sinh(\alpha t)^\frac{18}{\beta(n+2)}
\end{aligned}\right\}
$$
$$
\label{91f}\left.
\begin{aligned}\hskip 0.6cm
\varrho_8(t)=-6\Big((-3n^2-6n-9)\cosh(\alpha t)^2+(n+2)^2\beta\Big)\alpha^2\Big((-54\alpha^2(n^2+2n+3)\gamma\\& \hskip -12.6cm+\beta^2(n+2)^2)\cosh(\alpha t)^2-\beta(n+2)^2(-18\alpha^2\gamma+\beta)\Big)\sinh(\alpha t)^\frac{12}{\beta(n+2)}+(n+2)^2\beta^2\\& \hskip -12.6cm(\cosh(\alpha t)-1)(\cosh(\alpha t)+1)\Big(((-108\alpha ^2(n^2+2n+3)\gamma+\beta^2(n+2)^2)\cosh(\alpha t)^2\\& \hskip -12.6cm-\beta(n+2)^2(-36\alpha^2\gamma+\beta))\sinh(\alpha t)^\frac{6}{\beta(n+2)}-4\beta^2\gamma(\cosh(\alpha t)-1)(\cosh(\alpha t)+1)(n+2)^2)\Big)
\end{aligned}\right\}
$$
$$
\begin{aligned}\hskip -7.6cm
\label{g55}\phi_2(t)=\frac{\varrho_9(t)}{\varrho_6(t)}
\end{aligned}
$$
where
$$
\begin{aligned}
\label{g58}\varrho_9(t)&=\lambda18(n+2)^2\sinh(\alpha t)^\frac{6+(2n+4)\beta}{\beta(n+2)}\beta^2\Big(-\Big((n+2)^2\beta-3n^2-6n-9\Big)\alpha^2\cosh(\alpha t)^2\\& \hskip 0.6cm\sinh(\alpha t)^\frac{6}{\beta(n+2)}+(n+2)^2\beta\Big(\alpha^2\sinh(\alpha t)^\frac{6+(2n+4)\beta}{\beta(n+2)}+\frac{\beta\sinh(\alpha t)^2}{3}\Big)\Big)^2
\end{aligned}
$$
$$
\begin{aligned}\hskip -6.9cm
\label{g62}\phi_3(t)=\frac{\varrho_{10}}{\varrho_{13}}
\end{aligned}
$$
where
$$
\label{g59}
\begin{aligned}\hskip -2.9cm
\varrho_{10}(t)&=-2\Big(\varrho_{11}-\varrho_{12}+\beta^2(\cosh(\alpha t)-1)(\cosh(\alpha t)+1)(n+2)^2\Big)
\end{aligned}
$$
$$
\label{91i}\left.
\begin{aligned}\hskip 0.9cm
\varrho_{11}(t)=\Big( \Big(\mu^4(n+2)^6\beta^6-2916\alpha^6\gamma(n^2+2n+3)^3\Big)\cosh(\alpha t)^6)-3\beta(n+2)^2\Big(\mu^4(n+2)^4\beta^5\\& \hskip -14.6cm-972\alpha^6\gamma(n^2+n+3)^2\Big)\cosh(\alpha t)^4+3\Big(\mu^4(n+2)^4\beta^4-324\alpha^6\gamma(n^2+2n+3)\Big)\beta^2\\& \hskip -14.6cm(n+2)^4\cosh(\alpha t)^2-\beta^3(n+2)^6(-108\alpha^6 \gamma+\beta^3 \mu^4)\Big)\sinh(\alpha t)^\frac{18}{\beta(n+2)}
\end{aligned}\right\}
$$
$$
\label{91j}\left.
\begin{aligned}\hskip 0.9cm
\varrho_{12}(t)=108\beta^2 (n+2)^2 \gamma(\cosh(\alpha t)+1)\Big((-3n^2-6n-9)\cosh(\alpha t)^2+(n+2)^2\beta\Big)^2\alpha^4\\& \hskip -13.1cm(\cosh(\alpha t)-1)\sinh(\alpha t)^\frac{12}{\beta(n+2)}+36\beta^4(n+2)^4\gamma(\cosh(\alpha t)+1)^2\Big((-3n^2-6n-9)\\& \hskip -13.1cm\cosh(\alpha t)^2+(n+2)^2\beta\Big)\alpha^2(\cosh(\alpha t)-1)^2\sinh(\alpha t)^\frac{6}{\beta(n+2)}-4\gamma\beta^6(\cosh(\alpha t)-1)^3\\& \hskip -13.1cm(\cosh(\alpha t)+1)^3(n+2)^6\Big((-3n^2-6n-9)\cosh(\alpha t)^2+(n+2)^2\beta\Big)\alpha^2 \sinh(\alpha t)^\frac{6}{\beta(n+2)}
\end{aligned}\right\}
$$
$$
\label{91l}
\begin{aligned}\hskip -5.9cm
\varrho_{13}(t)=(\varrho_7-\varrho_{14})\sinh(\alpha t)^\frac{6}{\beta(n+2)}(n+2)^2\beta^2\sinh(\alpha t)^2
\end{aligned}
$$
$$
\label{91m}\left.
\begin{aligned}\hskip 0.9cm
\varrho_{14}(t)=-24\Big(\Big((n+2)^2\beta^2-54\alpha^2\gamma(n^2+2n+3)\Big)\cosh(\alpha t)^2-\beta(n+2)^2(-18\alpha^2\gamma+\beta)\Big)\\& \hskip -13.1cm(n+2)^2\alpha^2(\cosh(\alpha t)-1)\beta^2(\cosh(\alpha t)+1)((-3n^2-6n-9)\cosh(\alpha t)^2\\& \hskip -13.1cm+(n+2)^2\beta)\sinh(\alpha t)^\frac{12}{\beta(n+2)}+4(n+2)^4\Big(\Big((n+2)^2\beta^2-108\alpha^2\gamma(n^2+2n+3)\Big)\\& \hskip -13.1cm\cosh(\alpha t)^2-\beta(n+2)^2(-36\alpha^2\gamma+\beta)\Big)(\cosh(\alpha t)-1)^2\beta^4(\cosh(\alpha t)+1)^2\\& \hskip -13.1cm\sinh(\alpha t)^\frac{6}{\beta(n+2)}-16\gamma\beta^6(\cosh(\alpha t)-1)^3(\cosh(\alpha t)+1)^3(n+2)^6
\end{aligned}\right\}
$$
$$
\begin{aligned}\hskip -7.4cm
\label{g63}\phi_4(t)=\frac{-36\alpha^2\varrho_{15}\varrho_{16}}{(\varrho_7-\varrho_{14})\varrho_{17}}
\end{aligned}
$$
where
$$
\label{g71}\left.
\begin{aligned}\hskip 0.9cm
\varrho_{15}(t)&=\Big(\mu^4(n+2)^6\beta^6+5832\alpha^6\gamma(n^2+2n+3)^3\Big)\cosh(\alpha t)^6-3\beta(n+2)^2\Big(\mu^4(n+2)^4\beta^5\\& \hskip 0.6cm+1944\alpha^6\gamma(n^2+2n+3)^2\Big)\cosh(\alpha t)^4+3\beta^2(n+2)^4\Big(\mu^4(n+2)^2\beta^4+648\alpha^6\gamma\\& \hskip 0.6cm(n^2+2n+3)\Big)\cosh(\alpha t)^2-\beta^3(n+2)^6(216\alpha^6\gamma+\beta^3\mu^4)\Big)\sinh(\alpha t)^\frac{18}{\beta(n+2)}+\varrho_{25}
\end{aligned}\right\}
$$
$$
\label{91p}\left.
\begin{aligned}\hskip 0.9cm
\varrho_{25}(t)&=216(\cosh(\alpha t)+1)\beta^2(\cosh(\alpha t)-1)(n+2)^2\gamma\alpha^4\Big((-3n^2-6n-9)\cosh(\alpha t)^2\\& \hskip 0.6cm+\beta(n+2)^2\Big)^2\sinh(\alpha t)^\frac{12}{\beta(n+2)}-72(\cosh(\alpha t)+1)^2\beta^4(\cosh(\alpha t)-1)^2(n+2)^4\gamma\alpha^4\\& \hskip 0.6cm\Big((-3n^2-6n-9)\cosh(\alpha t)^2+\beta(n+2)^2\Big)\sinh(\alpha t)^\frac{6}{\beta(n+2)}+8\gamma\beta^6(\cosh(\alpha t)-1)^3\\& \hskip 0.6cm(\cosh(\alpha t)+1)^3(n+2)^6
\end{aligned}\right\}
$$
$$
\begin{aligned}\hskip 0.6cm
\label{g58i}\varrho_{16}(t)&=\Big((\beta(n+2)^2-3n^2-6n-9)\sinh(\alpha t)^\frac{6}{\beta(n+2)}-\beta(\cosh(\alpha t)-1)(\cosh(\alpha t)+1)\\& \hskip 9.6cm(n+2)\Big)\cosh(\alpha t)^2
\end{aligned}
$$
$$
\begin{aligned}\hskip 1.3cm
\label{g91j}\varrho_{17}(t)&=(-3\alpha^2((-3n^2-6n-9)\cosh(\alpha t)^2+(n+2)^2\beta)\sinh(\alpha t)^\frac{6}{\beta(n+2)}+\beta^2(\cosh(\alpha t)-1)\\& \hskip 7.6cm(\cosh(\alpha t)+1)(n+2)^2)\beta(n+2)\sinh(\alpha t)^2
\end{aligned}
$$
$$
\begin{aligned}
\label{g64}\hskip -7.1cm\phi_5(t)=\frac{24\varrho_{15}\varrho_{18}}{(\varrho_7-\varrho_{14})\varrho_{19}}
\end{aligned}
$$
where
$$
\begin{aligned}
\label{g58j}\hskip -1.1cm
\varrho_{18}(t)=\Big(((n+2)^2\beta-3n^2-6n-9)\alpha^2\Big(\cosh(\alpha t)^2+\frac{1}{2}\Big)\sinh(\alpha t)^\frac{6}{\beta(n+2)}\\-\frac{(\cosh(\alpha t)-1)(6\cosh(\alpha t)^2+\beta(n+2))(\cosh(\alpha t)+1)}{2}\Big)\alpha^2	
\end{aligned}
$$
$$
\begin{aligned}
\label{91k}\hskip -1.9cm
\varrho_{19}(t)=-3\alpha^2((-3n^2-6n-9)\cosh(\alpha t)^2+(n+2)^2\beta)\sinh(\alpha t)^\frac{6}{\beta(n+2)}\\+\beta^2(\cosh(\alpha t)-1)(\cosh(\alpha t)+1)(n+2)^2	
\end{aligned}
$$

$$
\begin{aligned}
\label{g65}\hskip -7.2cm
\phi_6(t)=\frac{\varrho_{20}}{\varrho_{21}\varrho_{22}}
\end{aligned}
$$
where
$$
\begin{aligned}
\label{g58k}\hskip 1.4cm
\varrho_{20}(t)=	-\Big(((n+2)^2\beta-n^2-6n-9)\alpha^2\sinh(\alpha t)^\frac{6+(2n+4)\beta}{\beta(n+2)}-((n+2)^2)-3n^2-6n-9)\cosh(\alpha t)^2\\\alpha^2\sinh(\alpha t)^\frac{6}{\beta(n+2)}+\beta\sinh(\alpha t)^2(n+2)\Big)^2\sinh(\alpha t)^\frac{18+(4n+8)\beta}{\beta(n+2)}\beta^6\cosh(\alpha t)^2(n+2)^6\alpha^2\mu^4
\end{aligned}
$$
$$
\begin{aligned}
\label{g58l}\hskip 1.2cm
\varrho_{21}(t)=
\Big(\alpha^2 \beta(n+2)^2\sinh(\alpha t)^\frac{6+(2n+4\beta)}{\beta(n+2)}-((n+2)^2\beta-3n^2-6n-9)\cosh(\alpha t)^2\alpha^2\sinh(\alpha t)^\frac{6}{\beta(n+2)}\\+\frac{\beta^2\sinh(\alpha t)^2(n+2)^2}{3}\Big)^2
\end{aligned}
$$
$$
\label{g60}
\begin{aligned}\hskip -4.4cm
\varrho_{22}(t)&=\varrho_{23}+\varrho_{24}(n+2)^2\beta^2(\cosh(\alpha t)-1)(\cosh(\alpha t)+1)
\end{aligned}
$$
$$
\label{91g}\left.
\begin{aligned}\hskip 0.9cm
\varrho_{23}(t)=\Big(\varrho_{26}+\frac{1}{8}((n+2)^2\beta(\mu^4(n+2)^4\beta^5+72\alpha^4(n^2+2n+3)(n+2)^2\beta^2\\& \hskip -11.6cm+108\alpha^4(n^2+2n+3)^2\beta-3888\alpha^6(n^2+2n+3)^2\gamma)\cosh(\alpha t)^4)-\frac{1}{8}(n+2)^4\beta^2\\& \hskip -11.6cm(\mu^4(n+2)^2\beta^4+12\alpha^4(n+2)^2\beta^2+72\alpha^4(n^2+2n+3)\beta-1296\alpha^6(n^2+2n+3)\\& \hskip -11.6cm\gamma)\cosh(\alpha t)^2+\frac{1}{24}\beta^3(n+2)^6(432\alpha^6\gamma+\mu^4\beta^3+36\alpha^4\beta)\Big)\sinh(\alpha t)^\frac{18}{\beta(n+2)}
\end{aligned}\right\}
$$
$$
\label{91q}
\begin{aligned}\hskip 0.9cm
\varrho_{26}(t)=\Big(\frac{-\mu^4(n+2)^6\beta^6}{24}-\frac{27\alpha^4(n+2)^2(n^2+2n+3)^2\beta^2}{2}+486\alpha^6(n^2+2n+3)^3\gamma\Big)\cosh(\alpha t)^6
\end{aligned}
$$
$$
\label{91h}\left.
\begin{aligned}\hskip 0.9cm
\varrho_{24}(t)=\Big(((-3n^2-6n-9)\cosh(\alpha t)^2+(n+2)^2\beta)\alpha^2((-54\alpha^2(n^2+2n+3)\gamma+\beta^2(n+2)^2)\\& \hskip -14.6cm\cosh(\alpha t)^2-\beta(n+2)^2(-18\alpha^2\gamma+\beta))\sinh(\alpha t)^\frac{12}{\beta(n+2)}+\frac{1}{3}\Big(2(n+2)^2\beta^2\Big(\Big((\frac{-\beta^2(n+2)^2}{4}\\& \hskip -14.6cm+27\alpha^2(n^2+2n+3)\gamma)\cosh(\alpha t)^2+\frac{1}{4}\beta(n+2)^2(-36\alpha^2\gamma+\beta)\Big)\sinh(\alpha t)^\frac{6}{\beta(n+2)}+\beta^2\gamma\\& \hskip -14.6cm(\cosh(\alpha t)-1)(\cosh(\alpha t)+1)(n+2)^2\Big)(\cosh(\alpha t)-1)(\cosh(\alpha t)+1)\Big)\Big)
\end{aligned}\right\}
$$
$$
\begin{aligned}
\label{g65i}\hskip -7.2cm
\phi_7(t)=\frac{-36n\alpha^2\varrho_{15}\varrho_{16}}{\varrho_{17}(\varrho_7-\varrho_{14})}
\end{aligned}
$$
\textbf{For case(ii)} The values of $\phi_1$, $\phi_2$, $\phi_3$, $\phi_4$, $\phi_5$, $\phi_6$ and $\phi_7$ are given below \\
$$
\begin{aligned}
\label{g29}\hskip -7.4cm
\phi_1(t)=\frac{48\delta_3(\pi+\frac{3\lambda}{16})}{\delta_1}\\
\end{aligned}
$$\\
where
$$
\begin{aligned}\label{g41}
\delta_1=\Big(\Big(\Big(\nu(n+2)^2\beta^2-18\alpha^2(n^2+2n+3)\Big)\cosh(\alpha t)^2-
\beta(n+2)^2(-6\alpha^2+\beta \nu)\Big)\\\sinh(\alpha t)^\frac{6}{\beta(n+2)}-2\beta^2(\cosh(\alpha t)-1)(\cosh(\alpha t)+1)(n+2)^2\Big)
\end{aligned}
$$\\
$$
\begin{aligned}
\label{g36}\hskip -1.9cm
\delta_3=\Big(\Big((-3n^2-6n-9)\cosh(\alpha t)^2+\beta(n+2)^2\Big)
\alpha^2 \sinh(\alpha t)^\frac{6}{\beta(n+2)}\\-\frac{\beta^2 \sinh(\alpha t)^2(\cosh(\alpha t)-1)(\cosh(\alpha t)+1)(n+2)^2}{3}\Big)
\end{aligned}
$$\\
$$
\begin{aligned}
\label{g35}\hskip -7.4cm
\phi_2(t)=\frac{-\delta_2\lambda}{\delta_1}
\end{aligned}
$$\\
where
$$
\begin{aligned}
\label{g42}\hskip -1.9cm
\delta_2=-3\Big((-3n^2-6n-9)\cosh(\alpha t)^2+\beta(n+2)^2\Big)\alpha^2\sinh(\alpha t)^\frac{6}{\beta(n+2)}\\+\beta^2(\cosh(\alpha t)-1)(\cosh(\alpha t)+1)(n+2)^2
\end{aligned}
$$\\
$$
\begin{aligned}
\label{g37}\hskip -7.4cm
\phi_3(t)=\frac{-3\nu\delta_3\delta_4}{\delta_1}
\end{aligned}
$$\\
where
$$
\begin{aligned}
\label{g43}\hskip -4.2cm
\delta_4=ln\Big(\frac{1}{\beta^2(n+2)^2\sinh(\alpha t)^2}\Big(\delta_9\Big)+ln(2)+ln(\tau)-1\Big)
\end{aligned}
$$\\
$$
\begin{aligned}
\label{91r}\hskip -1.9cm
\delta_9=-\beta^2(\cosh(\alpha t)-1)^2(\cosh(\alpha t)+1)^2(n+2)^2\sinh(\alpha t)^\frac{-6+(-2n-4)\beta}{\beta(n+2)}\\+3\alpha^2\Big((-3n^2-6n-9)\cosh(\alpha t)^2+\beta(n+2)^2\Big)
\end{aligned}
$$\\
$$
\begin{aligned}
\label{g38}\hskip -7.4cm
\phi_4(t)=\frac{\delta_5}{\delta_2\delta_1}
\end{aligned}
$$\\
where
$$
\begin{aligned}
\label{g74i}\hskip 0.9cm
\delta_5=18\Big(-\beta(\cosh(\alpha t)-1)(\cosh(\alpha t)+1)(n+2)\sinh(\alpha t)^\frac{6}{\beta(n+2)}+(\beta(n+2)^2\\-3n^2-6n-9)\sinh(\alpha t)^\frac{12}{\beta(n+2)}\alpha^2\Big)(n+2)\nu\beta\cosh(\alpha t)^2\alpha^2
\end{aligned}
$$
$$
\begin{aligned}
\label{g39}\hskip -7.4cm
\phi_5(t)=\frac{\delta_6}{\delta_2\delta_1}
\end{aligned}
$$\\
where
$$
\begin{aligned}
\label{g74j}\hskip 1.2cm
\delta_6=-12(n+2)^2\nu\beta^2\Big(\delta_{10}+(\beta(n+2)^2-3n^2-6n-9)(\cosh(\alpha t)^2+\frac{1}{2})\sinh(\alpha t)^\frac{12}{\beta(n+2)}\alpha^2\Big)\alpha^2
\end{aligned}
$$
$$
\begin{aligned}
\label{91s}\hskip -0.9cm
\delta_{10}=\frac{-(\cosh(\alpha t)-1)(\cosh(\alpha t)+1)(6\cosh(\alpha t)^2+\beta(n+2))\sinh(\alpha t)^\frac{6}{\beta(n+2)}}{2}
\end{aligned}
$$
$$
\begin{aligned}
\label{g40}\hskip -7.4cm
\phi_6(t)=\frac{\delta_7}{\delta_8\delta_1}
\end{aligned}
$$\\
where
$$
\begin{aligned}
\label{g44}\hskip 1.2cm
\delta_7=4(n+2)^2\sinh(\alpha t)^\frac{6}{\beta(n+2)}\nu\beta^2\cosh(\alpha t)^2\Big((\beta(n+2)^2-3n^2-6n-9)\alpha^2]\sinh(\alpha t)^\frac{6}{\beta(n+2)}\\-\beta(\cosh(\alpha t)-1)(\cosh(\alpha t)+1)(n+2)\Big)^2\alpha^2
\end{aligned}
$$
$$
\begin{aligned}
\label{g44i}\hskip -1.9cm
\delta_8=\Big(-(\beta(n+2)^2-3n^2-6n-9)\cosh(\alpha t)^2\alpha^2\sinh(\alpha t)^\frac{6}{\beta(n+2)}+\delta_{11}\Big)^2
\end{aligned}
$$
$$
\begin{aligned}
\label{91t}\hskip -4.9cm
\delta_{11}=(n+2)^2\beta\Big(\alpha^2\sinh(\alpha t)^\frac{6+(2n+4)\beta}{\beta(n+2)}+\frac{\beta\sinh(\alpha t)^2}{3}\Big)
\end{aligned}
$$

$$
\begin{aligned}
\label{g40i}\hskip -7.4cm
\phi_7(t)=\frac{n\delta_5}{\delta_2\delta_1}
\end{aligned}
$$\\
\textbf{For case(iii)} The values of $\phi_1$, $\phi_2$, $\phi_3$, $\phi_4$, $\phi_5$, $\phi_6$ and $\phi_7$ are given below \\
$$
\begin{aligned}\hskip -7.5cm
\label{g45}\phi_1(t)=\frac{-16\pi-3\lambda}{2\zeta_1-2}
\end{aligned}
$$
where
$$
\begin{aligned}\hskip -6.2cm
\label{g51}\zeta_1=\kappa \iota e^\frac{-2\iota\Big(\zeta_8-3\alpha^2\Big((-3n^2-6n-9)\cosh(\alpha t)^2+\beta(n+2)^2\Big)\Big)}{\beta^2(n+2)^2\sinh(\alpha t)^2}
\end{aligned}
$$
$$
\begin{aligned}\hskip -2.9cm
\label{91n}\zeta_8=\beta^2(\cosh(\alpha t)-1)^2(\cosh(\alpha t)+1)^2(n+2)^2\sinh(\alpha t)^\frac{-6+(-2n-4)\beta}{\beta(n+2)}
\end{aligned}
$$
$$
\begin{aligned}
\label{g46}\hskip -7.5cm\hskip -0.2cm\phi_2(t)=-\frac{\lambda}{2\zeta_1-2}
\end{aligned}
$$
$$
\begin{aligned}
\label{g47}\hskip -7.5cm\phi_3(t)=\frac{\zeta_2\zeta_4}{\zeta_3}
\end{aligned}
$$
where
$$
\begin{aligned}
\label{g52}\hskip -2.9cm\zeta_2=e^\frac{2\Big(\beta^2\sinh(\alpha t)^2(\cosh(\alpha t)^4+1)(n+2)^2\sinh(\alpha t)^\frac{-6+(-4n-8)\beta}{\beta(n+2)}+9\alpha^2\cosh(\alpha t)^2(n^2+2n+3)\Big)\iota}{\beta^2(n+2)^2\sinh(\alpha t)^2}
\end{aligned}
$$
$$
\begin{aligned}
\label{g53}\hskip -0.1cm\zeta_3=\beta^2(n+2)^2\sinh(\alpha t)^4\Big(\iota \kappa e^\frac{2\iota\Big(\beta^2(\cosh(\alpha t)^4+1)(n+2)^2\sinh(\alpha t)\frac{-6+(-2n-4)\beta}{\beta(n+2)}+9\alpha^2\cosh(\alpha t)^2(n^2+2n+3)\Big)}{\beta^2(n+2)^2\sinh(\alpha t)^2} \\\hskip -4.6cm-e^\frac{4\iota\Big(\sinh(\alpha t)\frac{-6+(-2n-4)\beta}{\beta(n+2)}\cosh(\alpha t)^2\beta+\frac{3\alpha^2}{2}\Big)}{\beta\sinh(\alpha t)^2}\Big)
\end{aligned}
$$
$$
\begin{aligned}
\label{g53i}\hskip -0.6cm\zeta_4=\kappa\Big(\beta^2\iota(\cosh(\alpha t)-1)(\cosh(\alpha t)+1)(n+2)^2\sinh(\alpha t)^\frac{-6}{\beta(n+2)}+\Big(\frac{-(n+2)^2\beta^2}{2}\\+9\alpha^2\iota(n^2+2n+3)\Big)\cosh(\alpha t)^2-3(n+2)^2\beta(\iota\alpha^2-\frac{\beta}{6})\Big)
\end{aligned}
$$
$$
\begin{aligned}
\label{g48}\hskip -7.5cm\phi_4(t)=\frac{-36\alpha^2\zeta_5\zeta_2\iota^2\kappa}{\beta(n+2)\zeta_3}
\end{aligned}
$$
where
$$
\begin{aligned}
\label{g53j}\hskip -1.2cm\zeta_5=\beta(\cosh(\alpha t)-1)(\cosh(\alpha t)+1)(n+2)\sinh(\alpha t)^\frac{-6}{\beta(n+2)}+\alpha^2\Big((n+2)^2\\\beta-3n^2-6n-9\Big)\cosh(\alpha t)^2
\end{aligned}
$$
$$
\begin{aligned}
\label{g49}\hskip -7.5cm\phi_5(t)=\frac{-24\iota^2\zeta_2\alpha^2\kappa\zeta_6}{\zeta_3}
\end{aligned}
$$
where
$$
\begin{aligned}
\label{g53k}\hskip -3.1cm\zeta_6=-\zeta_8+\Big(\cosh(\alpha t)^2+\frac{1}{2}\Big)((n+2)^2\beta-3n^2-6n-9)\alpha^2
\end{aligned}
$$
$$
\begin{aligned}
\label{91u}\hskip -0.9cm\zeta_8=\frac{(\cosh(\alpha t)-1)(6\cosh(\alpha t)^2+\beta(n+2))(\cosh(\alpha t)+1)\sinh(\alpha t)^\frac{-6}{\beta(n+2)}}{2}
\end{aligned}
$$
$$
\begin{aligned}
\label{g50}\hskip -7.5cm\phi_6(t)=\frac{-144\alpha^2\iota^2\cosh(\alpha t)^2\zeta_1\zeta_7}{(n+2)^4\beta^4(\zeta_1-1)}
\end{aligned}
$$
$$
\begin{aligned}
\label{g50i}\hskip -7.5cm\phi_7(t)=\frac{-36n\alpha^2\zeta_5\zeta_2\iota^2\kappa}{\beta(n+2)\zeta_3}
\end{aligned}
$$
$$
\begin{aligned}
\label{g53l}\hskip -1.4cm\zeta_7=\Big(\alpha^2((n+2)^2\beta-3n^2-6n-9)\sinh(\alpha t)^\frac{6}{\beta(n+2)}-\beta(\cosh(\alpha t)-1)\\(\cosh(\alpha t)+1)(n+2)\Big)^2\sinh(\alpha t)^\frac{-12+(-6n-12)\beta}{\beta(n+2)}
\end{aligned}
$$

\end{document}